\documentclass[conference]{IEEEtran}
\IEEEoverridecommandlockouts
\usepackage{cite}
\usepackage{amsmath,amssymb,amsfonts}
\usepackage{algorithmic}
\usepackage{graphicx}
\usepackage{textcomp}
\usepackage{xcolor}
\usepackage{balance}
\usepackage{setspace}
\usepackage{graphicx,rotating,booktabs}
\usepackage{multirow}
\usepackage{tabularx}
\usepackage{blindtext}
\usepackage{color}
\usepackage{graphics}
\usepackage{listings}
\usepackage{color}
\usepackage{tabulary} 
\usepackage{subcaption}
\usepackage[skip=2pt]{caption}
\definecolor{dkgreen}{rgb}{0,0.6,0}
\definecolor{gray}{rgb}{0.5,0.5,0.5}
\definecolor{mauve}{rgb}{0.58,0,0.82}

\def\BibTeX{{\rm B\kern-.05em{\sc i\kern-.025em b}\kern-.08em
    T\kern-.1667em\lower.7ex\hbox{E}\kern-.125emX}}

\newcommand{\algocc}{{\sc{\small SourcererCC}}}
\newcommand{\algodp}{{\sc{\small AutoenCODE}}}
\begin{document}

\title{Clone Detection on Large Scala Codebases}

%{\footnotesize \textsuperscript{*}Note: Sub-titles are not captured in Xplore and
%should not be used}
%\thanks{Identify applicable funding agency here. If none, delete this.}
%}

% \author{\IEEEauthorblockN{1\textsuperscript{st} Given Name Surname}
% \IEEEauthorblockA{\textit{dept. name of organization (of Aff.)} \\
% \textit{name of organization (of Aff.)}\\
% City, Country \\
% email address or ORCID}
% \and
% \IEEEauthorblockN{2\textsuperscript{nd} Given Name Surname}
% \IEEEauthorblockA{\textit{dept. name of organization (of Aff.)} \\
% \textit{name of organization (of Aff.)}\\
% City, Country \\
% email address or ORCID}
% }

\author{
\IEEEauthorblockN{Wahidur Rahman\IEEEauthorrefmark{1,3}, Yisen Xu\IEEEauthorrefmark{2}, Fan Pu\IEEEauthorrefmark{2}, Jifeng Xuan\IEEEauthorrefmark{2}, Xiangyang Jia\IEEEauthorrefmark{2},\\Michail Basios\IEEEauthorrefmark{3}, Leslie Kanthan\IEEEauthorrefmark{3}, Lingbo Li\IEEEauthorrefmark{3}, Fan Wu\IEEEauthorrefmark{3} and Baowen Xu\IEEEauthorrefmark{4}}
\IEEEauthorblockA{\IEEEauthorrefmark{1}Imperial College London, London, United Kingdom\\w.rahman17@imperial.ac.uk}
\IEEEauthorblockA{\IEEEauthorrefmark{2}Wuhan University, Wuhan, China\\\{xuyisen,fan\_pu,jxuan,jxy\}@whu.edu.cn}
\IEEEauthorblockA{\IEEEauthorrefmark{3}Turing Intelligence Technology, London, United Kingdom\\\{mike, leslie, lingbo, fan\}@turintech.ai}
\IEEEauthorblockA{\IEEEauthorrefmark{4}Nanjing University, Nanjing, China\\bwxu@nju.edu.cn}}

% \author{\IEEEauthorblockN{Wahidur Rahman}
% \IEEEauthorblockA{Imperial College London, London\\
% United Kingdom\\
% w.rahman17@imperial.ac.uk}
% \and
% \IEEEauthorblockN{Yisen Xu, Fan Pu, Jifeng Xuan, Xiangyang Jia}
% \IEEEauthorblockA{Wuhan University, Wuhan\\
% China\\
% \{xuyisen,fan\_pu,jxuan,jxy\}@whu.edu.cn}
% \and
% \IEEEauthorblockN{Michail Basios, Leslie Kanthan, Lingbo Li, Fan Wu}
% \IEEEauthorblockA{Turing Intelligence Technology, London\\
% United Kingdom\\
% \{mike, leslie, lingbo, fan\}@turintech.ai}
% \and
% \IEEEauthorblockN{Baowen Xu}
% \IEEEauthorblockA{Nanjing University, Nanjing\\
% China\\
% bwxu@nju.edu.cn}
% }

\maketitle

\begin{abstract}
Code clones are identical or similar code segments. The wide existence of code clones can increase the cost of maintenance and jeopardise the quality of software. 
The research community has developed many techniques to detect code clones, however, there is little evidence of how these techniques may perform in industrial use cases.
In this paper, we aim to uncover the differences when such techniques are applied in industrial use cases.
We conducted large scale experimental research on the performance of two state-of-the-art code clone detection techniques, \emph{SourcererCC} and \emph{AutoenCODE}, on both open source projects and an industrial project written in the Scala language.
Our results reveal that both algorithms perform differently on the industrial project, with the largest drop in precision being 30.7\%, and the largest increase in recall being 32.4\%.
By manually labelling samples of the industrial project by its developers, we discovered that there are substantially less Type-3 clones in the aforementioned project than that in the open source projects.
%Upon discussion with industrial developers, we discovered that developers tend to consider refactorability when judging Type-3 clones since unrefactorable clones have little value to them.
%Therefore, refactorability should be taken into account during the detection process for clone detection techniques to be more practical in industrial use case.
\end{abstract}

%
% Keywords. The author(s) should pick words that accurately describe the work being
% presented. Separate the keywords with commas.

\begin{spacing}{0.94}

\section{Introduction}
\label{sec_introduction}
With more and more software source code constantly being published on open source platforms such as \emph{GitHub}, code reuses or code clones are common to many repositories ~\cite{7962379,Gabel:2010:SUS:1882291.1882315}.
Code clones are segments of source code that are identical or similar in syntax or semantics~\cite{RATTAN20131165, Sheneamersurvey}.
Developers often create code clones via copying existing code and pasting with or without modifications to reduce the development time and costs.
However, code clones were shown to increase software maintenance costs, since software bugs can easily propagate via code cloning and inconsistent fixing of these bugs may induce undefined behaviour~\cite{5070547}.
While completely avoiding code clones is impractical, researchers have proposed different detection techniques to identify code clones in existing codebases to reduce the risk of code clones.

Existing clone detection techniques can be  generally categorised according to the code representation used in the algorithm: text-based~\cite{792593}, lexical- or token-based~\cite{1019480, 7886988}, graph-based~\cite{10.1007/3-540-47764-0_3}, abstract syntax tree-based~\cite{4023995, Jiang:2007:DSA:1248820.1248843}, bytecode-based~\cite{6240495, 8338420}, or their combinations~\cite{Funaro:2010:HAC:1808901.1808914, 5609665}.
Among different clone detection techniques, token-based algorithms have shown promising results whilst being scalable to large  datasets~\cite{Ragkhitwetsagul2018}. 
One state-of-the-art token-based algorithm is \algocc{}~\cite{7886988}, which incorporates several optimisation techniques that dramatically improve its speed. 
Meanwhile, with growing attention to deep learning, White et al. and Tufano et al. introduced and improved a deep learning technique, \algodp{} for clone detection~\cite{7582748, 8595238}, which outperforms many previous algorithms.

Despite promising reports, the performance of these state-of-the-art algorithms have only been applied to open source codebases, where code reuse and foraging is a common practice~\cite{7962379}. 
In contrast, to protect their source code from leaking, many commercial organisations allow limited or no access to public open source platforms. 
This makes code reuse from external sources difficult or even impossible. 
In such a closed programming environment, detected code clones are likely to be different from those in an open environment where there are rich sources for code cloning.
Therefore, investigation is required for ascertaining whether the performance of clone detection algorithms may change when applied to closed codebases.

In this paper, we empirically investigate the performance of clone detection on an industrial project with over $4$ million lines of code (LoC) written in Scala, a functional programming language.
We aim to examine two state-of-the-art clone detection algorithms, namely \algocc{} and \algodp{}, on a private industrial project and determine the potential differences in performance when applying them to open source projects.
Both algorithms are adapted to accept programs written in Scala as the industrial codebase is written in Scala.
The source code in the industrial codebase was created in an environment where downloading code from external sources was strictly forbidden.
Such programming environment is not uncommon in many industrial organisations, therefore, the private industrial codebase we studied in this paper can be an example of how clone detection techniques may perform in industrial use cases in general.
To compare the performance of the algorithms on industrial and open source codebases, we apply the algorithms to the industrial codebase and the top $20$ Scala projects on \emph{Github}, ranked according to the number of stars of the projects.
Samples of code clones are manually labelled by human developers, and precision and recall metrics are calculated as measurements of the algorithms' performance.

%By measuring the performance of the clone detection algorithms on open source Scala projects, we learn that the precision is comparable with the performance reported by the original authors.
%It demonstrates that both algorithms remain unaffected by the shift to functional languages when precision is concerned. 
%The \algodp{} technique however, exhibits a drop of up to 28.2\% in recall. 
Experimental results show that, when applied to the industrial codebase, both algorithms show different degrees of degradation in precision, with the biggest drop from 95.1\% to 64.4\%, but the recall can improve as much as 32.4\%.
We also observed a substantially lower proportion of Type-3 clones (definitions of different types are introduced in Section~\ref{sec_clone_detection}) in the industrial codebase, and a higher proportion of the samples are not considered as clones.
Upon discussion with industrial developers, we discovered that many developers may consider unrefactorable clones as having little or no value to them and may be reluctant to classify such code segments as clones.

\section{Background}
\label{sec_clone_detection}

\subsection{Scala Language}
Scala is a general purpose programming language that makes use of both, object-oriented and functional programming paradigms. It is closely related to the Java programming language as it compiles to java bytecode and provides interoperability with packages written in Java, but also provides functional programming features such as currying, higher order functions, type inference and pattern matching. 
While evaluation of clone detection techniques on the Java language exists in great number in the literature, evaluation on Scala or other functional languages is lacking.

\vspace{-5pt}
\begin{lstlisting}[caption={Example of two similar methods in Scala language},label={lst_clone_example},language=Scala,basicstyle=\footnotesize]
  def secondElementIfArray(x: Any) = x match {
    case Array(_, a, _*) => a
    case _ => "default"
  }
  
  def nameIfDog(x: Any) = x match {
    case Dog(a) => a
    case _ => "default"
  }
\end{lstlisting}

An example of a syntactically similar method pair in Scala language is shown in Listing~\ref{lst_clone_example}. 
In the example, the first method checks whether the input is an array with at least two elements and returns the second element of the array.
The second method checks whether the input is of the class \texttt{Dog}, where \texttt{Dog(name: String)} is the constructor of the class.
It returns its member variable \texttt{Dog.name} if it is the matching class.
Though these two methods implement quite different functionalities, due to the \emph{pattern matching} feature of Scala, the two methods appear similar in syntax. Therefore, these two methods can be identified as clones by clone detection algorithms.
However, clone detection for Scala programs may be distinct from that for other languages. 
For instance, if the two methods in Listing~\ref{lst_clone_example} are implemented in Java, one may use size checking on an array and an array elements' access, while the other requires class type checking and the \emph{getter} function of a member variable.
Therefore, they may appear different and subsequently make a clone detection algorithm identify them as non-clones. 

\subsection{Types of Code Clones}
We follow the widely accepted definitions of different types of clones~\cite{RATTAN20131165,Sheneamersurvey,7886988}:
%A brief overview of the types of code clones is provided here with examples written in scala. These clone types can be broadly categorized into four different groups as follows:

\textbf{Type-1}. Identical code fragments except for differences in whitespace, comments and layout.

\textbf{Type-2}. Identical code fragments except for differences in identifier names and literal values, in addition to Type-1 clone differences.

\textbf{Type-3}. Syntactically similar code fragments that contain added, modified and/or removed code statements with respect to each other, in addition to Type-1 and Type-2 clone differences.

\textbf{Type-4}. Syntactically different code fragments, but are semantically similar in terms of their implemented functionalities.

\subsection{\algocc{} and \algodp{}}
\label{sec_scc}

In this paper, we investigate the performance of two state-of-the-art clone detection techniques; \algocc{} is a token-based algorithm and \algodp{} incorporates deep learning with different code representations \cite{7886988, 8595238}.

\algocc{} is a state-of-the-art token-based clone detection algorithm proposed by Sajnani et al.~\cite{7886988}.
It represents a code fragment as a bag of tokens (tokens that appear in the fragment and their frequencies).
The clone detection criterion is deterministic and based on the degree of overlap between the bags of tokens from two code fragments, and a percentage similarity threshold is used as a cutoff when deciding whether they are clones. 
This simple idea requires pairwise comparisons of all methods to assess the degree of overlap and is therefore $O(n^2)$ complexity and hard to scale to large real-world projects.
\algocc{} overcomes this by exploiting two properties of token features, which define the upper and lower bounds of similarity if only a subset of tokens is seen.
%This algorithm operates in two primary stages: (i) partial index creation; and (ii) clone detection.
Through the creation of a memory efficient partial index for candidates that satisfy these properties, and repetitively updating the upper and lower bounds of similarity measure as more tokens are seen, it is then able to eliminate the majority of method pairs as early as possible with certainty. 
With such optimisation, the number of method pairs requiring a full comparison is greatly reduced, thus the speed of the algorithm is dramatically increased.

\algodp{} is a state-of-the-art clone detection algorithm based on deep learning, a neural network based technique to minimise the need for manual feature engineering. 
It works by generating sentence embeddings using deep learning on four different representations of code fragments: \emph{Identifier}, Abstract Syntax Tree (\emph{AST}), compiled bytecode, and Control Flow Graph (\emph{CFG}). 
The first two representations can be obtained from the source code, and the latter two are usually obtained from compiled binary code.
\algodp{} works primarily in four stages.
In the first stage, code representations are extracted from both source code and binary code, and word vectors are generated at the required code granularity level (in our case, we require a vector for each method).
In the second stage, word embeddings for each word in the word vector of a method are learnt using a Recurrent Neural Network.
In the third stage a sentence embedding is learnt for each method with a Recursive Auto Encoder~\cite{DBLP:conf/emnlp/SocherPHNM11}, using the word embeddings from the second stage.
In the final stage, euclidean distances between sentence embeddings are computed and a distance threshold is used to determine which methods are clones.
The algorithm detects clones using each code representation independently, and the results can be combined in different ways to form the final result.

\section{Experiment Design}
\label{sec_experiment_design}
In this section, we describe in details the datasets, the experiment procedure, and the Research Questions (RQs).

\subsection{Datasets}
\label{sec_datasets}
In order to evaluate and compare the performance of clone detection algorithms on open source and industrial codebases, we use two separate datasets in our experiments: an open source dataset and a private industrial dataset.

The open source dataset is composed of the top 20 most popular Scala projects (that received the most stars) on \emph{GitHub}.
Since more popular projects tend to have more developers contributing to the project, the collaborative coding practice should be the closest to that of an industrial project.
A summary of the 20 Scala projects, including their lines of code and number of methods, is outlined in Table~\ref{table_open_source_results}.
The numbers of lines of code are counted after comments and empty lines are removed, and only the methods with a minimum of 10 lines are considered, which is a common practice for clone detection~\cite{RATTAN20131165,7886988}.
The project sizes vary from $2,091$ to $305,276$ lines of code, and the numbers of methods vary from $30$ to $5,256$.
%The varying sizes help in understanding the scalability of the algorithms at different degree of project sizes.

The industrial project is obtained via a collaboration with an industry organisation, which wishes to remain anonymous.
The project consists of more than 4 million of lines of code, and is written and maintained by hundreds of developers within the company.
Due to the specific domain of the company, developers are given limited access to open source platforms. For instance, open platforms such as \emph{SourceForge} and \emph{BitBucket} are completely blocked within the company. 
\emph{GitHub} can be viewed, but downloading open source projects without approval is strictly forbidden. 
Similar requirements can be found in many other private companies to protect the proprietary source code~\cite{BosuCBOC17,SalamK18}.
Code written in such an environment is to our interests as it is difficult or even impossible to copy open source contents and the code cloning behaviours can be different.
This may lead to a different pattern of code clones in the codebase, and therefore different performance of clone detection algorithms.
To gain access to this private codebase, some of the authors worked within the company for a period of time to conduct the experiments.

\begin{table}[t]
\centering
\caption{Projects summary.}
\resizebox{0.45\textwidth}{!}{
\begin{tabular}{lrr}
\toprule
\label{table_open_source_results}
Project  & Method Count & Lines of Code   \\ \hline
scala-2.13.x          & 5,256 & 305,276   \\ 
playframework         & 894  & 69,653  \\ 
gitbucket             & 371  & 22,754  \\ 
finagle               & 1,663 & 130,195  \\ 
kafka-manager         & 238  &  13,310  \\ 
lila                  & 1,161 & 77,669  \\ 
bfg-repo-cleaner      & 30  & 2,091  \\ 
fpinscala             & 67   & 7,378   \\ 
gatling               & 432  & 33,375 \\ 
scalaz-series-7.3.x   & 435  & 44,237  \\ 
incubator-openwhisk   & 591 & 52,938   \\ 
sbt                   & 865  & 44,350  \\ 
scala-js              & 3,373 & 133,952  \\ 
scala-native          & 1,610 & 103,170  \\ 
dotty                 & 4,310 & 284,081  \\ 
scalding              & 745  & 48,440  \\ 
BigDL                 & 2,376 & 163,881  \\ 
breeze                & 1,244 & 48,271  \\ 
shapeless             & 604 & 30,491  \\ 
spray                 & 380  &  33,167  \\ 
\textbf{Open Source Total} & 26,645 & 1,648,679 \\
\textbf{Industrial Project}            & 69,533  & 4,051,596  \\ \bottomrule
                              
\end{tabular}}
\end{table}

\subsection{Experiment Procedure}
\label{sec_experiment_procedure}
%Describe the experiments in details. How we removed the cycle relations in clones, and how we sampled the data.

Our experiments consist of three steps: data filtering, data labelling, and clone detection for both open source projects and the industrial project. We describe the details of each step as follows.

\subsubsection{Data Filtering}

\label{sec_data_filtering}

% In order to evaluate the performance of clone detection algorithms, the ground truth of clone pairs has to be obtained first. 
Instead of relying on other clone detection tools as oracle~\cite{oreo}, we choose to manually label the data as our ground truth.
The benefit of manual labelling is not just better accuracy, but also, for the industrial dataset, the developer labelled data can better represent the industrial viewpoint on what Type-3 and Type-4 clones are.
As the number of method pairs grows quadratically with the number of methods, it is impractical to label all method pairs.
%For instance, with almost 70k methods in the industrial codebase, there are more than 4 billion method pairs, the vast majority of which are actually not clones.
Therefore, we use \algocc{} to filter method pairs such that we can focus on the pairs that are more likely to be clones.
A 70\% similarity threshold was given to the \algocc{} algorithm in the data filtering stage.
Such threshold was chosen according to our observation that method pairs below this threshold are very rarely to be clones, therefore can be regarded as non-clones without affecting our results.
%The algorithm was reported to perform better with a threshold of over 90\%.
%By using a threshold of 70\%, we ensure not only potential clone pairs, but also some ``almost'' clone pairs are included for the next stage of data labelling.
%
% After filtering by \algocc{}, the remaining potential clone pairs could form clone groups, where methods in a group can be clones to each other.
% This is undesirable not only because they introduce redundancy in the data, but also introduce bias to the results as bigger groups of similar methods weigh more in the results.
% Therefore, we conducted a simple reduction algorithm to remove such redundancy.
% If each method is a node in a graph, and a potential clone pair forms an edge between the two methods in the pair, the redundancy can be represented by the cycles in the graph.
% Therefore, we iteratively remove a random edge in a cycle until there is no cycle in the graph.
%As a result, the remaining potential clone pairs form a forest, or an undirected acyclic graph.
%For the industrial dataset for example, after \algocc{} filtering, the number of method pairs is reduced to 19,634.

\subsubsection{Data Labelling}

For different datasets, similar approaches are used for data labelling.
For the industrial dataset, we asked the developers who have contributed to the codebase to voluntarily label the data.
We sent out the invitation of labelling to a group of 705 developers, and 67 developers responded to the invitation by labelling at least one of the method pairs. 
To make the labelling process simple and convenient, we created a web-based GUI within the company's system to show a randomly picked method pair to the participant, along with the definition of the four different types of clones and their corresponding examples. 
The participant is asked to label the method pair as which type of clone, or not a clone, based on his/her best knowledge. 
%Each method pair has an equal chance to be presented until at least two developers agree on the labelling.
Owing to the timing and the voluntary nature of the data labelling, we obtained 201 labelled method pairs that at least two developers agreed on the label.
%By the time of writing, there are 862 method pairs that are labelled by at least one developer, among which 201 pairs reach consensus (at least two developers agree).
%These 201 pairs are then used to evaluate the performance of the algorithms.

For the open source dataset, two of the authors labelled the data separately, when there was disagreement, another author would make a decision between the two different labels.
All of the three authors who labelled the open source dataset had been working intensively on the adaptation of the clone detection algorithms to Scala language, therefore, all three authors understood Scala programs and labelled the data with confidence.

\subsubsection{Clone Detection}

To evaluate the performance of \algocc{} and \algodp{}, we run both algorithms on the industrial codebase and open source codebases.
In this paper, we only use two code representation for \algodp{}: \emph{Identifier} (leaf) and \emph{AST} (path), as opposed to using all four representations in the original publication~\cite{7886988}.
This is because the other two representations require the binary code after compilation, which is forbidden for the industrial codebase; meanwhile, our preliminary experiments showed that the other two representations did not work well for Scala language.
%For the open source codebases, we run the algorithms on each project separately.
%This is because we are interested in the scalability of the algorithms across different sizes of codebases, and because the industrial codebase emphasises intra-project clones (as explained in Section~\ref{sec_datasets}, cloning from sources outside the company is difficult), thus running the algorithms on each open source project separately to have comparable results.
The results of both industrial and open source datasets are then validated by the labelled data. We calculated precision and recall metrics to compare the performance of the algorithms.

\subsection{Experimental Setup}
%Algorithm parameters, running machine specifications.
For \algocc{} on both industrial and open source datasets, a 90\% similarity threshold is used to identify code clones.
For \algodp{}, we keep the default parameters recommended by the original authors.
%When running \algodp{} on the industrial dataset, due to the size of the project, we had to slightly adjust the parameters such that the algorithm can finish within 24 hours on the allocated machine in the company.
%Specifically, we keep all the parameters the same, but adjust \textit{maximum\_number\_of\_iterations} parameter from 30 to 2.
%This is because the use case for the company is to run the algorithm on a weekly basis, such that they can review the identified code clones and take actions accordingly.
%With such a large codebase contributed by hundreds of developers, there is a large number of commits to the codebase each day.
%Therefore, identifying code clones in a version that is more than one day old is much less valuable to the developers.
%Therefore, the adjusted parameter is determined by the algorithm being able to finish on the industrial codebase within 24 hours.
%Our preliminary study illustrates that there is no statistically significant difference in the results from such change of parameters on open source projects.
%Moreover, we consider any potential loss of performance due to the adjustment of parameters on the industrial project as part of the difference between the performance on industrial codebases and open source codebases that we are investigating in this paper, since many algorithms cannot be run with their optimal parameters when applied on real-world problems, due to other requirements or restrictions such as timing and data availability.

The machine specifications for running the clone detection algorithms differ between the open source and industrial projects due to the company regulations regarding the use and location of the source code. 
Experiments on the industrial codebase were run on a virtual machine with six cores of Intel Xeon CPU and 64GB of memory;
Experiments on the open source projects were run on Google Cloud instances with eight cores of virtual CPU and 30GB of memory.

\textbf{Implementation}.
We re-implemented variants of algorithms \algocc{} and \algodp{} to support Scala language. 
%Currently, the \algocc{} tool does not have have support for Scala, 
For \algocc{}, we re-implemented the algorithm fully in Scala with the native \textit{scala.meta} library for parsing Scala code.
%Adapting the technique to work with Scala after tokenization is relatively straightforward as we can follow the steps of Sajnani et al. to implement the algorithm.
For \algodp{}, we also use \textit{scala.meta} library to parse Scala code and extract tokens for the \emph{Identifier} and \emph{AST} representations.
For the rest of the algorithm, we use the implementation from the original authors.
% Compiled binary code is used to extract Bytecode and CFG representations.
Our implementation of \algocc{} and \algodp{} can be found on \emph{GitHub}\footnote{https://GitHub.com/Wahidur-Rahman/scala\_sourcererCC}\footnote{https://GitHub.com/Wahidur-Rahman/scala\_autoencode}.

\subsection{Research Questions}
\label{sec_research_questions}
In this paper, we aim to answer the following RQs. 
We explained the rationals behind each question as follows.

\begin{itemize}
\item[\textbf{RQ1}] What is the performance of \algocc{} and \algodp{} on open source codebases after adapted for Scala Language?
\end{itemize}

This RQ is to understand the baseline of the performance of clone detection algorithms under investigation.
Specifically, we are interested in whether an algorithm performs as good on Scala language as they do with other languages studied by previous researchers. 
We answer this question by measuring precision and recall metrics of both algorithms on 20 open source Scala projects. 
The execution time of the algorithms on each project is recorded to evaluate the scalability of the algorithms.

% \begin{itemize}
% \item[\textbf{RQ2}] How is the quality of the data labelling by industrial developers?
% \end{itemize}

% Since we are relying on human developers to label the industrial data, 
% we first want to understand whether the labelling is of a similar quality as of academic researchers.
% To answer this question, we count how many cases the first two developers reached consensus.
% Intuitively, the more often unanimous agreements occur, the more consistent the views of developers are, and therefore the more reliable the labels are.
% Similarly, the same measures are obtained for the labelling of academic researchers  on open source projects as a comparison.

\begin{itemize}
\item[\textbf{RQ2}] Is there any difference in the performance of the clone detection algorithms when they are applied on open source codebase and private industrial codebase?
\end{itemize}

We pose this RQ to understand whether the state-of-the-art clone detection algorithms perform differently on private industrial codebase.
Using the data labelled by human developers, we calculate the precision and recall metrics of both algorithms, and compare them with their respective performances on open source datasets.

\section{Results}
\label{sec_results}
%In this section, we present the results and answer the research questions in completeness. 

\subsection{RQ1. Results on Open Source Benchmarks}
The first research question queries the performance of \algocc{} and \algodp{} on open source Scala projects.
We assess the results of both algorithms on the 20 open source benchmarks listed in Table~\ref{table_open_source_results}.

To evaluate the performance of the algorithms, we manually labelled 1000 random samples of method pairs after the data filtering process.
Using the labelled data and the result from both clone detection algorithms, we can draw the Confusion matrix to understand its performance (Table~\ref{confusion_scala}). 
From the table, we can calculate \textit{precision} and \textit{recall} measurements of the algorithms.
For instance, the \textit{precision} and \textit{recall} of \algocc{} are calculated as:
$$ precision = \frac{TP}{TP+FP} = \frac{247}{247+1} = 99.6\%$$
$$ recall = \frac{TP}{TP+FN} = \frac{247}{247+616} = 28.6\% $$
where $TP$, $FP$, $FN$ are the counts of True Positive, False Positive, and False Negative respectively.

\begin{table*}[htbp]
\small
\centering
\caption{Confusion matrices of \algocc{} and \algodp{} on the open source projects and the Industrial project. The \emph{combination} columns are formed using the union of the Identifier and AST representation clones from the \algodp{} algorithm.}
%Results of \algodp{} are presented by \emph{Identifier} representation, \emph{AST} representation, and combination of both representations, respectively.}
\label{confusion_scala}
\resizebox{0.9\textwidth}{!}{
\begin{subtable}{.55\linewidth}
\begin{tabular}{@{\extracolsep{4pt}}l rrr}
\toprule
\multicolumn{2}{l}{\multirow{2}{*}{Open Source projects}} & \multicolumn{2}{l}{Truth (From Authors)}\\ 
\cmidrule{3-4} 
\multicolumn{2}{l}{}          & \textbf{Clone}          & \textbf{Not Clone}\\ \hline
\multirow{2}{*}{\algocc{} identification}            & \textbf{Clone}    & 247            & 1\\ 
 & \textbf{Not Clone} & 616              & 136 \\
\cmidrule{1-4} 
\multirow{2}{*}{\algodp{} \emph{(Identifier)}}            & \textbf{Clone}    & 57            & 1 \\ 
 & \textbf{Not Clone} & 806              & 136  \\
\cmidrule{1-4} 
\multirow{2}{*}{\algodp{} \emph{(AST)}}            & \textbf{Clone}    & 117            & 6 \\
 & \textbf{Not Clone} & 746              & 131  \\
\cmidrule{1-4} 
\multirow{2}{*}{\algodp{} (Combination)}            & \textbf{Clone}    & 139            & 7 \\ 
 & \textbf{Not Clone} & 724              & 130  \\

\bottomrule

\end{tabular}
\end{subtable}
\begin{subtable}{.55\linewidth}
\begin{tabular}{lrrr}
\toprule
\multicolumn{2}{l}{\multirow{2}{*}{Industrial Project}} & \multicolumn{2}{l}{Truth (From Developers)}\\ \cmidrule{3-4}
\multicolumn{2}{l}{}          & \textbf{Clone}          & \textbf{Not Clone}\\ \hline
\multirow{2}{*}{\algocc{} identification}            & \textbf{Clone}    & 35            & 10\\  
 & \textbf{Not Clone} & 28              & 128 \\
\cmidrule{1-4}
\multirow{2}{*}{\algodp{} \emph{(Identifier)}}            & \textbf{Clone}    & 13            & 2 \\ 
 & \textbf{Not Clone} & 50              & 136  \\
\cmidrule{1-4}
\multirow{2}{*}{\algodp{} \emph{(AST)}}            & \textbf{Clone}    & 29            & 16 \\  
 & \textbf{Not Clone} & 34              & 122  \\
\cmidrule{1-4}
\multirow{2}{*}{\algodp{} \emph{(Combination)}}            & \textbf{Clone}    & 30            & 16 \\  
 & \textbf{Not Clone} & 33              & 122  \\
\bottomrule
\end{tabular}
\end{subtable}
}
\end{table*}

\begin{figure}[ht]
    \centering
    \includegraphics[width=0.9\linewidth]{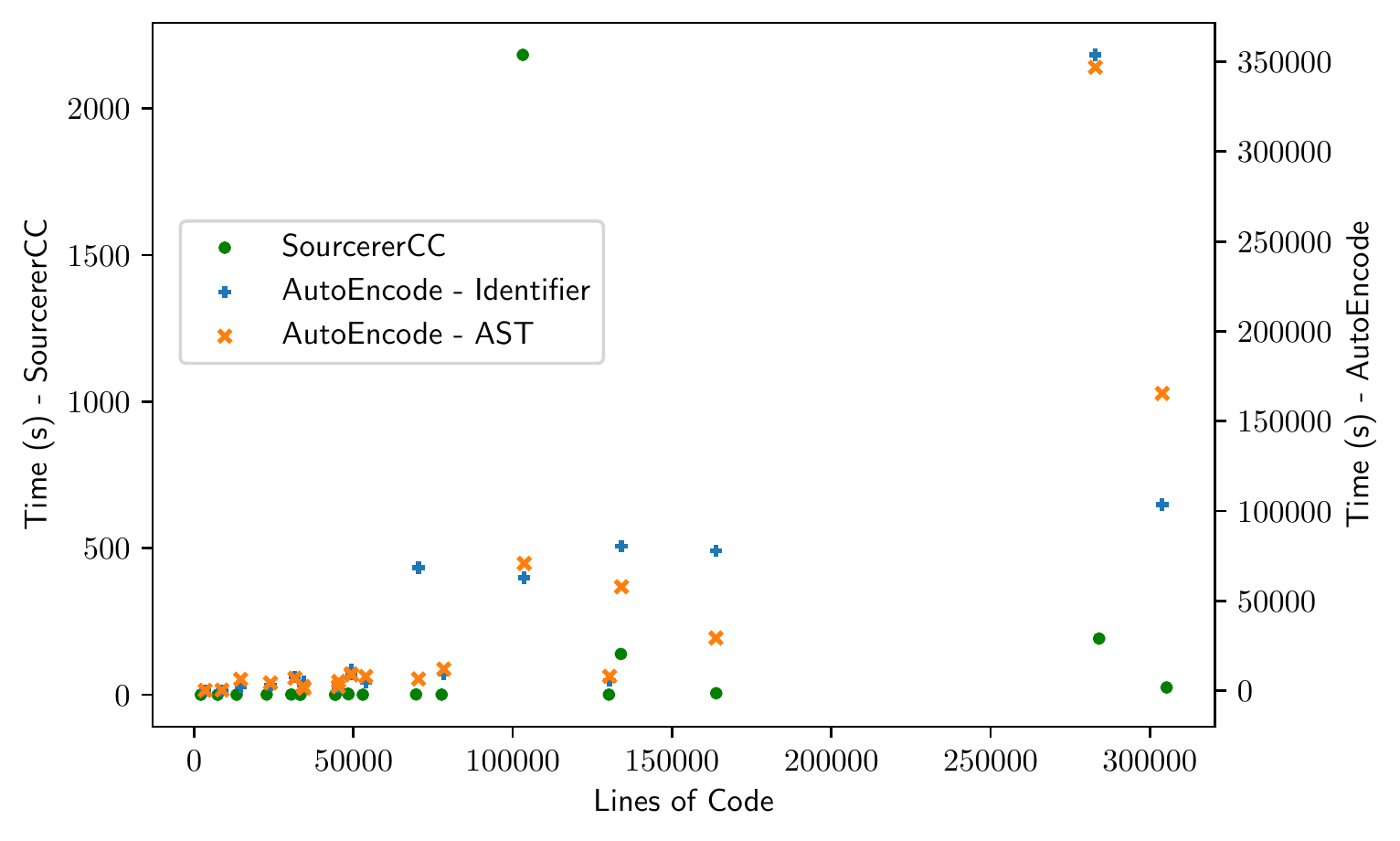}{}
    \caption{Execution time as the Lines of Code increases for the open source benchmarks}
    \label{fig_timing_loc}
\end{figure}

\begin{figure}[ht]
    \centering
    \includegraphics[width=0.9\linewidth]{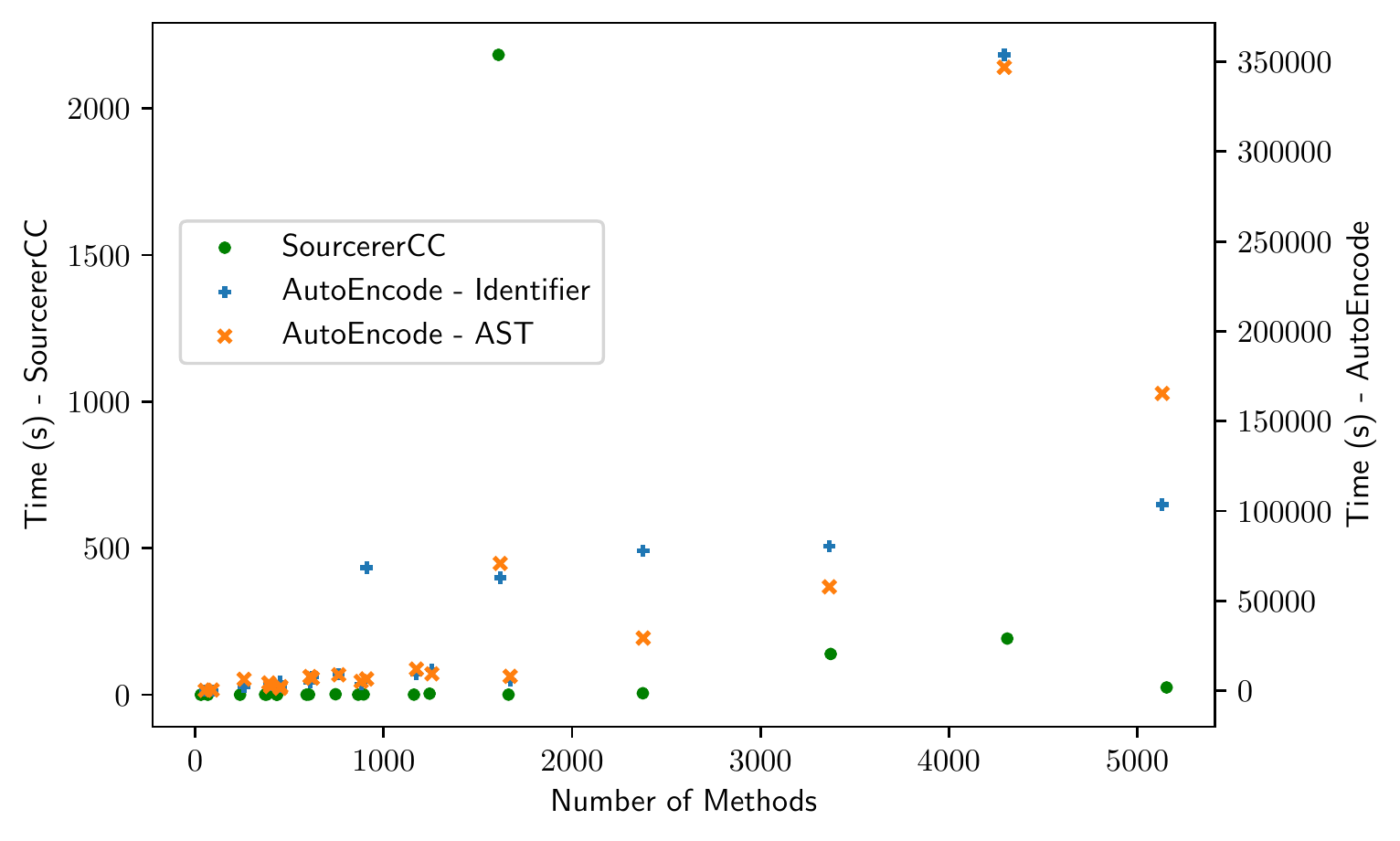}{}
    \caption{Execution time as the Number of Methods increases for the open source benchmarks}
    \label{fig_timing_nom}
\end{figure}

Precision and recall values are calculated for each algorithm, and are summarised in Table~\ref{table_precision_recall_scala}.
According to the table, we can see that both algorithms show a high level of precision, with \algocc{} topping the table with 99.6\% of precision. 
The precision of the \algocc{} algorithm is in keeping with that seen by Sajnani et al.~\cite{7886988}, where they calculated a precision of 91\% on a sample of clones from \textit{BigCloneBench}~\cite{Svajlenko:2015:ECD:2881297.2881379}. 
For the \algodp{} algorithm, the precision values are as well similar to that reported by Tufano et al.~\cite{8595238}.
According to their results, \algodp{} achieved 100\% and 96\% precision for \emph{Identifier} and \emph{AST} representations respectively, whereas it is 98.3\% and 95.1\% respectively in our findings.
The recall values cannot be compared with previous studies directly since we used different means to obtain the "ground truth" of False Negatives.

we plot the execution time for all open source projects against their lines of code and number of methods in Figure~\ref{fig_timing_loc} and Figure~\ref{fig_timing_nom} respectively.
The execution time for each algorithm uses a different scale on the y axis in the Figures. 

The execution time of \algocc{} falls mostly in the range of 0.087 to 192 seconds, with an outlier of 2183 seconds observed on project \emph{scala-native}. On the other hand, the execution time of \algodp{} is generally 3 degrees of magnitude larger, ranging from 4.5 to 5781 minutes. 
These numbers are consistent with the execution time reported by both algorithms' original authors, where \algocc{} took a few seconds for projects in the order of $10^6$ lines of code, and \algodp{} could take several hours.

\textbf{Summary}. Both \algocc{} and \algodp{} achieve similar precision measurements on open source Scala projects, indicating that both algorithms retain performance when applied to functional languages. Though we use our own implementation of both algorithms, the execution time of both algorithms is consistent with their original authors. The timing of algorithm \algodp{} can be up to 3 degrees of magnitude longer than \algocc{}.

\begin{table}[!t]
\small
\caption{Precision and recall metrics for \algodp{} and \algocc{} on the open source and industrial projects}
\label{table_precision_recall_scala}
\centering
\resizebox{0.46\textwidth}{!}{
\begin{tabular}{@{\extracolsep{4pt}}lrrrr}
\toprule
& \multicolumn{2}{c}{Open Source projects} & \multicolumn{2}{c}{Industrial project} \\ \cmidrule{4-5} \cmidrule{2-3} 
                                          & Precision            & Recall           & Precision          & Recall          \\ \midrule
\algocc{} & 99.6\% & 28.6\% & 77.8\% & 55.6\% \\ 
\algodp{} (\emph{Identifier}) & 98.3\% & 6.6\% & 86.7\% & 20.6\%  \\ 
\algodp{} (\emph{AST}) & 95.1\% & 13.6\% & 64.4\% & 46.0\% \\ 
\algodp{} (Combination) & 95.2\% & 16.1\% & 65.2\% & 47.6\% \\ \bottomrule
\end{tabular}
}
\end{table}

\subsection{RQ2. Performance Comparisons}
In this section, we assess the performance of the two algorithms on the industrial project, in terms of precision and recall metrics, and compare with their performance on open source projects. 
A summary of precision and recall values for both the industrial project and open source project can be found in Table~\ref{table_precision_recall_scala}.

For the \algocc{} algorithm, the precision measurements drop on the industrial project, compared with the performance on Scala open source projects, however, the recall measurements increase.
Precision drops from 99.6\% to 77.8\%, a difference of 21.8\%, but recall increases from 28.6\% to 55.6\%, a difference of 27\%.
For the \algodp{} algorithm, similar results can be observed.
Precision measurements drop for all representations, with the biggest drop observed from 95.1\% to 64.4\% on \emph{AST} representation, a difference of 30.7\%.
On the other hand, recall measurements increase for all representations, with the biggest increase from 13.6\% to 46\% on \emph{AST} representation, a difference of 32.4\%.
 
\textbf{Summary}. When both algorithms are applied to the industrial project, their performance changed substantially, with decreased precision and increased recall. The precision can drop as much as 30.7\%, while the recall may increase as much as 32.4\%.

\subsection{Discussion}
\label{sec_label_quality}

In order to understand why the performance of the algorithm is different on the industrial project, we summarise the distribution of different types of clones in Table~\ref{table_type_distribution}.

It is noticeable that there are differences in the distribution of the types of clones sampled from the open source project and the industrial project.
Firstly, Type-2 clones constitute 11.8\% of the samples from open source projects, while this number is only 7.5\% for the industrial project.
This can be due to the closed programming environment where the industrial project was created and maintained.
Due to the limited access to other code sources, direct copying and pasting code segments happens less often, which may contribute to the less prevalence of Type-2 clones.

\begin{table}[htbp]
\small
\centering
\caption{Distribution of different types of clones}
\label{table_type_distribution}
\resizebox{0.45\textwidth}{!}{
    \begin{tabular}{lrrrrrr}
    \toprule
    Open Source projects & Type 1 & Type 2 & Type 3 & Type 4 & Not a Clone & Total \\ \midrule
    Open Source projects & 53 & 118 & 641 & 51 & 137 & 1000 \\
    Open Source projects (\%) & 5.3 & 11.8 & 64.1 & 5.1 & 13.7 & 100 \\
    Industrial project & 11 & 15 & 28 & 9 & 138 & 201 \\
    Industrial project (\%) & 5.5 & 7.5 & 13.9 & 4.5 & 68.7 & 100 \\\bottomrule
    \end{tabular}
}
\end{table}

However, the differences in clone detection performance are more likely to be the direct result of much less Type-3 clones and much more non-clones in the industrial project sample. 
According to Table~\ref{table_type_distribution}, only 13.9\% of the industrial samples are Type-3 clones, while it is 64.1\% in open source samples. 
On the other hand, 68.7\% of the industrial samples are not a clone, while this number is only 13.7\% in open source samples.
Upon discussion with some developers who labelled the industrial samples, we discovered that their views of what should be a clone might be different from the academic standards.
For industrial use cases, developers tend to consider the goal of such clone detection is to help them refactor or remove duplicate code.
With that in mind, and when the judgement of a potential Type-3 clone can be subjective, industry developers tend to judge by whether it can be refactored, thus much less Type-3 clones.
This results directly in more False Positives and less False Negatives for the clone detection algorithms, therefore poorer precision and better recall observed in the industrial project.

\textbf{Summary}. The distributions of different types of clones are different for the labelled samples from open source projects and industrial project. Industrial developers tend to take refactorability into account for the judgement of potential Type-3 clones. Therefore, clone detection algorithms should take such factors into account to be more practical in industrial use case.

\section{Threats to Validity}
\label{sec_threats}

%\subsection{internal validity}

\textbf{Internal Validity}.  
The authors of \algocc{} used 6 lines/50 tokens as the minimum size of methods to be considered, whereas we used 10 lines in our experiments. 
We chose 10 lines because it is a common cutoff used in the majority of clone detection research, including the \algodp{} paper.
Furthermore, smaller methods are likely to include more uninteresting or trivial clones that are easier to detect~\cite{Gabel:2010:SUS:1882291.1882315}.
Therefore, if we were to include methods with 6-10 lines, the performance of \algocc{} measured in the context of our experiments is likely to be slightly better than that reported in this paper, thus it would still be comparable with the results from the original authors.

%The parameters with which the \algodp{} is run on the open source projects and the industrial project are slightly different.
%This may cause the algorithm to perform worse on the industrial projects.
%We mitigate this threat by conducting preliminary experiments with these two sets of parameters on the open source projects, which showed no statistically significant difference in the results.
% Furthermore, A similar degree of performance degradation is also seen in \algocc{}, which was run with the same parameters on the open source and the industrial projects.
% Therefore, if the \algodp{} algorithm is run with the same parameters and a less severe performance degradation may be found, the conclusion that clone detection techniques may suffer different degrees of performance degradation in industrial use cases, would still stand.

%\subsection{external validity}

\textbf{External Validity}. 
The performance of the algorithms were evaluated on 20 open source Scala projects from \emph{Github}, it may not be the same beyond these 20 projects.
We mitigate this threat by selecting the most popular Scala projects on \emph{Github}, measured by number of stars.
These projects are more likely to be forked or referenced by Scala program developers, such that the clone patterns may be carried on to many other Scala projects.
Therefore, the performance measured on these 20 projects should be representative for that on most Scala projects.

The performance differences seen in the industrial project studied in this paper may not generalise to other industrial projects.
However, this industrial project was developed in an environment where access to external code sources was very limited, which is a common practice in many companies in order to protect their proprietary source code.
Code created in such an environment is likely to have a similar cloning pattern.
Therefore, the evaluation on the industrial project in this paper should be representative for other industrial projects, and the threat is thus mitigated.

\section{Related Work}
\label{sec_related_work}

Code clone detection is widely studied. In this paper, we aim at studying the code clones in Scala programs and focusing on the difference from existing empirical results. We list related work as follows. 

Sheneamer et al.~\cite{Sheneamersurvey} surveyed different types of clone detection techniques up to 2016.
They categorised clone detection techniques into: textual approaches, lexical approaches, syntactical approaches, and semantic approaches, where the approaches in each category are more complicated than the category immediately preceding it. %he means to say lexical approaches is more complicated that textual approaches for instance
The performance of the algorithms were extracted from other references, whereby different datasets were used, but none was based on a private and closed codebase.  
Sajnani et al.~\cite{7886988} proposed \algocc{}, a token-based clone detection algorithm.
Despite the principle of the algorithm being simple, the authors applied several optimisations to eliminate impossible clone pairs as early as possible, and used an inverted index to make the algorithm scalable to very large codebases.
White et al.~\cite{7582748} and Tufano et al.~\cite{8595238} proposed and improved a Deep Learning-based clone detection algorithm that automatically learned features from four different representation of the source code at different levels of abstraction. 
The proposed algorithm identified potential code clones by calculating the similarities between code segments using those learned features.
Saini et al.~\cite{oreo} used a pipeline framework to identify code clones, where each stage of the pipeline used varied metrics of the code segments.
Their results revealed that the framework had better detection of Type-3 and Type-4 clones, which were generally difficult to detect for other clone detection algorithms.
Despite the aforementioned researchers having evaluated different clone detection algorithms on datasets as large as $100$ millions of lines of code, none have yet to evaluate the performance on private industrial codebase.

Bellon et al.~\cite{4288192} and Roy et al.~\cite{Roy:2009:CEC:1530898.1531101} compared the performance of different clone detection algorithms and tools up to 2007 and 2009.
To make an unbiased comparison, the algorithms and tools were applied on the same datasets, which were relatively small (less than one million lines of code).
Svajlenko et al.~\cite{6976098} extended the previous comparison framework for clone detection algorithms and tools, and introduced their mutation and injection framework.
Ragkhitwetsagul et al.~\cite{Ragkhitwetsagul2018} compared clone detection techniques as well as plagiarism detection and compression tools up to 2018, in a scenario by scenario basis.
When different algorithms and tools were compared with each other over a common dataset, it posed some requirements on the dataset, including programming language and size of the dataset.
As a result, all of the tool comparison works were done over some open source codebases.
Svajlenko et al.~\cite{7816515} introduced \textbf{BigCloneEval}, a clone detection tool evaluation framework, which makes it easier for new clone detection algorithms to be evaluated over a common open source dataset.
Regarding functional languages, Xu et al.~\cite{xu2019mining} investigated how developers used the unique features of Scala language.

There are also studies regarding code clones on industrial codebases.
Monden et al.~\cite{1011328} looked at the correlation between code clones and software quality in a quantitative way.
They used a token-based algorithm to identify code clones in an industrial codebase consisting of more than one million lines of code, but the performance of the algorithm was not evaluated, as it was not the focus of the paper.
Zhang et al.~\cite{6405284} investigated the reasons behind code cloning from the perspective of developers and organisations, by analysing code clones and interviewing developers.
They also applied a token-based algorithm to detect code clones in a large industrial codebase consisting of more than 14 millions lines of code.
A number of sampled code clones were manually inspected, but the performance of the algorithm was not reported, as it was not the focus of their paper.

\section{Conclusion}
\label{sec_conclusion}

In this paper, we revisited the performance of two state-of-the-art clone detection algorithms, \algocc{} and \algodp{}, in their adaptation to Scala. Our experiments are conducted on an industrial project and 20 open source projects. 
We found that both \algocc{} and \algodp{} retained consistent performance on open source Scala projects, when precision is concerned.
However, we observed that the precision dropped substantially when they are applied on the industrial project.

Further investigation shows that there are less Type-2 clones in the industrial project, which could be caused by the limited access to other code sources, and there are substantially less Type-3 clones in the industrial project.
Our initial discussion with industrial developers suggests that they consider unrefactorable clones unfavorable in the clone detection results, and they tend to classify such code segments as non-clones.
Such observation motivates further investigation on the cloning practices and adaptation of clone detection algorithms in industrial use cases.
%Additionally, our results show a clear difference in industrial developers' and academic researchers' views towards what is a clone, as when asked to label clone types, industrial developers are more likely to agree on Type-4 clones and non-clones, whereas academic researchers are more likely to agree on Type-1 and Type-2 clones.
%Such difference may also be a reason of the observed degradation in performance, thus clone detection techniques should take industrial perspective on code clones into account in industrial use cases.

In the future, we would like to extend this study to investigate in details what are considered clones from industrial point of view, and how we could adapt such view to guide Clone Detection algorithms to be more practical in industrial cases. Furthermore, we want to demonstrate that Clone Detection techniques can be used in various software optimisation studies~\cite{10.1145/3236024.3236043, 10.1007/978-3-319-66299-2-14, 10.1145/2739480.2754648}, as additional restriction or even optimisation objective during the optimisation process.

\section*{Acknowledgement}
The work is supported by the National Key R\&D Program of China under Grant No. 2018YFB1003901, and the National Natural Science Foundation of China under Grant No. 61872273.
%the Open Research Fund Program of CETC Key Laboratory of Aerospace Information Applications under Grant No. SXX18629T022, 
%and the Advance Research Projects of Civil Aerospace Technology, Intelligent Distribution Technology of Domestic Satellite Information, under Grant No. B0301.

\end{spacing}

\balance
\bibliographystyle{./IEEEtran}
%\bibliography{./bibliography/IEEEabrv,./bibliography/citations}
\bibliography{./main}
\end{document}